**Efficient Facemask Sterilization via Forced Ozone Convection**

*Joseph Schwan[1†], Troy R. Alva[3†], Giorgio Nava[1,2], Carla Berrospe Rodriguez[1,2], Justin W. Chartron[3], Joshua Morgan[3], Lorenzo Mangolini[1,2]\**

__________

[1]Department of Mechanical Engineering, University of California Riverside, 900 University Ave, Riverside, CA, 92521.

[2]Department of Material Science, University of California Riverside, 900 University Ave, Riverside, CA, 92521.

[3]Department of Bioengineering, University of California Riverside, 900 University Ave, Riverside, CA, 92521.

[†]These authors contributed equally.
*Email: lmangolini@engr.ucr.edu
__________

During the beginning of 2020, the Covid-19 pandemic took the world by surprise, rapidly spreading undetected between and within many countries and wreaking havoc on the global economy both through death tolls and lockdowns. Healthcare professionals treating the coronavirus patients grapple with a massive and unprecedented shortage of Facepiece Respirators (FPRs) and other personal protective equipment (PPE), which act as fundamental tools to protect the health of the medical staff treating the patients affected by the coronavirus.

As of as of May 18$^{th}$, 2020, this shortage has directly and indirectly lead to the deaths of over 1000 medical personnel, and as the looming threat of a second wave approaches the medical field braces for another shortage. While many FPRs are designed to be disposable single-use devices, the development of affordable and efficient sterilization strategies is necessary to circumvent future shortages. Here, we describe the development of a plasma-based method to sterilize PPE such as FPRs with ozone. The system's novel design uses a flow-through configuration where ozone directly flows through the fibers of the PPE through the maintenance of a pressure gradient. Canonical ozone-based methods place the mask into a sealed ozone-containing enclosure but lack pressurization to permeate the mask fibers. In this device, ozone is created through an atmospheric pressure Dielectric Barrier Discharge (DBD) fed with compressed air. Due to limited supply and clinical need of FPRs, we demonstrated sterilization with surgical masks. We demonstrate rapid sterilization using *E. coli* as a model pathogen. This decision was made based on the safety and availability of this pathogen, with understanding that bacterial cell walls make them generally harder to eradicate than viral pathogens. A flow-through configuration enables a >400% improvement of the sterilization efficiency with respect to the canonical approach. This method has potential for a broad and cost-effective utilization. Using the power supply from a readily available plasma ball toy, a plastic box, a glass tube, steel mesh, and 3D printed components, we designed and tested an extremely affordable portable prototype system for rapid single mask sterilization which produced comparable results to its large high-cost equivalent.

## 1 Introduction

In the beginning of 2020, COVID-19 rapidly emerged as a global pandemic that has resulted in hundreds of thousands of deaths. Unprepared for this crisis, healthcare professionals experienced a shortage of disposable Personal Protective Equipment (PPE); in particular Facepiece Respirators (FPRs), such as designated N95 masks in the US and the FFP3 respirators in Europe. These respirators are fundamental tools that protect medical personnel caring for COVID-19 patients. Their designations are earned by the ability to filter out 95% and 99% of particulate matter at or above 0.3 microns in size[1], the scale of an average virion. The response to this disease has been severely compromised by the lack of adequate PPE. Disruptions to the PPE global supply chain have led to month-long delivery times and massive price increases, leaving doctors and nurses unprotected. As manufacturers are called upon to meet demand, healthcare providers have improvised with less effective substitutes.[2] While, based on manufacturer recommendations, the FPRs are single-use PPE and the US Centers for Disease Control and Prevention (CDC) does not formally recommend their decontamination and re-use, it is acknowledged that in these times of scarcity, decontamination might be considered as a good "practical" solution.[3] The development of standardized approaches to decontaminate fibers, restore filtering electrostatic charge, and in general re-use FPRs is a necessary to mitigate impact on both humans and the environment due to their future increased use, as the World Health Organization estimated that a 40% increase of the global PPE supplies will be needed.[4]

Ozone ($O_3$) is an allotropic form of oxygen with proven pan-viricidal and bactericidal capabilities. It is already widely employed on an industrial scale for wastewater treatment.[5] Notably, $O_3$ has been reported as effective in de-activating other members of the coronavirus family[6,7] and the bacteriophage MS2,[8] a virus previously shown to be more resistant to UV-based disinfection with respect to coronaviruses.[9] Additionally, $O_3$ can be directly produced from air (e.g. via plasmas or irradiation with UV light) and reconverted into non-hazardous $O_2$ with the aid of catalytic converters.[10] Therefore, unlike other compounds, $O_3$ can be readily manufactured with cost-effective approaches at the point-of-use. As a gaseous sterilization agent it is a particularly promising option for disinfecting poorly accessible spaces within porous materials, such as FRPs. While both consumer-grade and large-scale $O_3$ sterilization devices are widely available for deodorizing and sanitizing both rooms and objects, the design of these systems is not optimized for the disinfection of FPRs. In consumer grade $O_3$ sterilization devices objects are loaded into a sterilization chamber which is then sealed and flooded with $O_3$. $O_3$ passively diffuses into the objects and may slowly enter the porous media of an FPR.

With this contribution we propose an efficient $O_3$ sterilization approach specifically designed for FPRs. Compressed air is fed into a cylindrical atmospheric pressure Dielectric Barrier Discharge (DBD) plasma that when ignited rapidly produces $O_3$. The gas flow is then forced directly through the porous media of the FPR, which is directly connected to the plasma reactor. This method uses a non-thermal plasma to produce $O_3$, which avoids thermal decomposition of FPRs as the output gas is near room temperature. The efficacy of this method was compared to the canonical method by quantifying the sterilization efficiency of surgical masks saturated with *E. coli*. This pathogen was chosen due to its safety, availability, and ease of use with recognition that the cell envelope presented by bacterial pathogens are generally more problematic to ozone than viral capsids, though there is still debate as to whether viruses are more susceptible overall.[11] Finally, we demonstrate that this approach can be readily adapted as a low-cost solution by using the power supply of a widely available commercial plasma globe toy, a few 3D printed parts, some steel mesh, and a plastic box, to construct a portable low-power system capable of attaining similar sterilization efficiencies.

## 2 Experimental Section

### 2.1 DBD Plasma reactor for FPR sterilization

**Figure 1** shows the DBD reactor used for the mask sterilization experiments. The system comprises a quartz tube (10 mm outer diameter; quartz wall thickness 1 mm), an outer copper electrode connected to a DC power supply (Trek High Voltage Amplifier 10/40A/HS connected to a signal generator; 10 kHz sinusoidal wave with amplitude between 1 kV and 10kV) and a 6 mm stainless steel tube as a grounded electrode. Compressed air is flown through the system at constant rate of 10 slm controlled by a King Instruments flowmeter.

The waveform of the discharge voltage $V$ was measured from the output of the power supply while the waveform of the discharge charge $Q$ was recorded using a 20 nF capacitor, serially connected to the grounded electrode. Both $V$ and $Q$ were recorded using a digital oscilloscope (Tektronix AFG320).

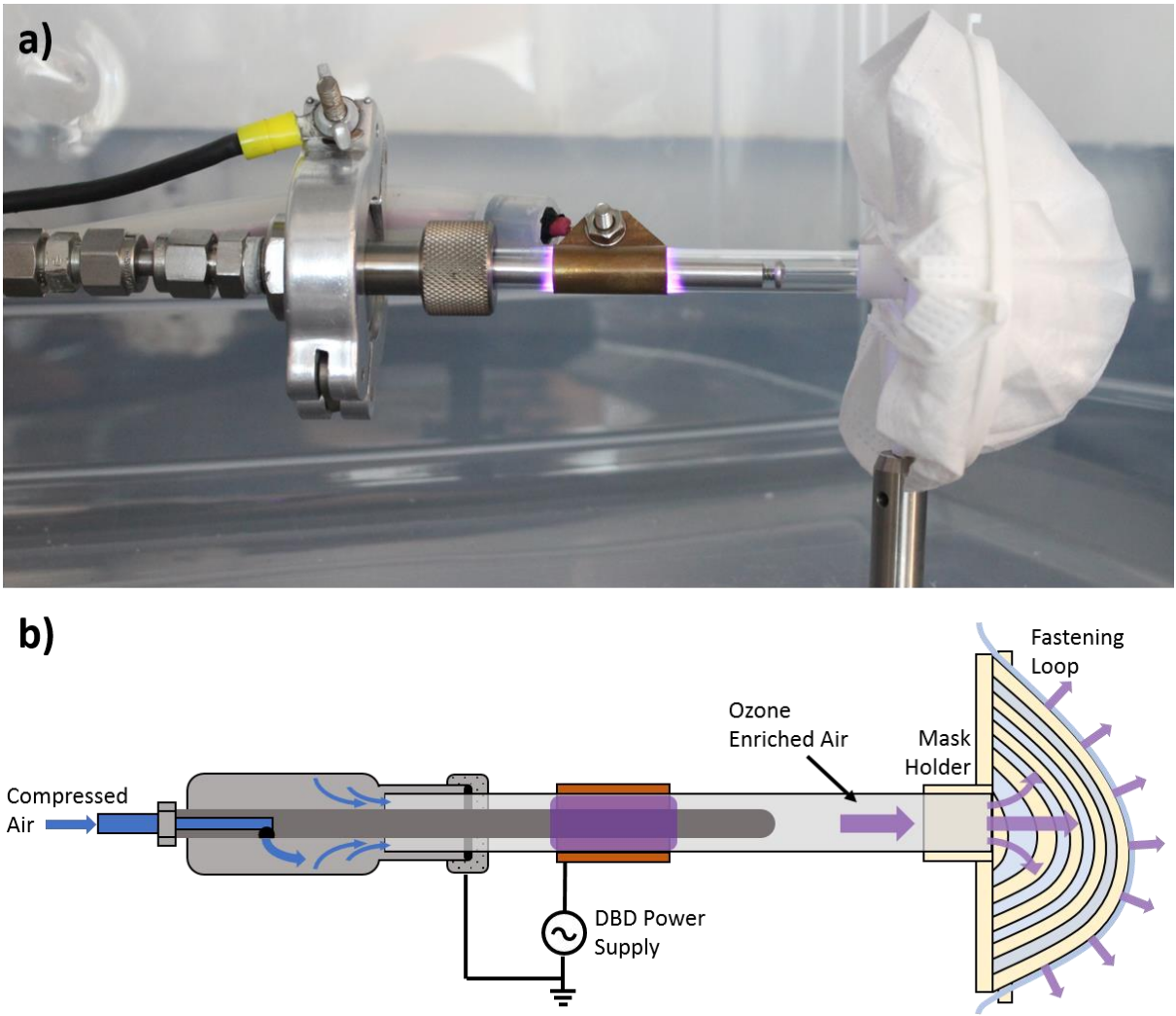

*Figure 1.* (a) Picture and (b) schematic of the DBD reactor used in the mask sterilizaton experiments.

## 2.2 Low-cost plasma reactor for FPR sterilization

Modifying a plasma globe toy (in these experiments a 6-inch Theefun plasma globe) can produce a system that is able to evolve $O_3$. This is due to the flyback transformer and the timer circuit within the system that produces low-current and high-voltage (1 kV to 6 kV) sawtooth or ramp output signals near or at a frequency of 30 kHz. This output is low enough in frequency and high enough in amplitude that it is capable of sustaining a DBD plasma. In order to take advantage of this system the electrode geometry was adjusted so that a steel mesh inside the reactor was powered and the external electrode (now consisting of metallic HVAC tape to further lower material costs) was grounded. Additional material was placed within the mesh acting as a flow control, forcing all compressed air to pass through the plasma region. **Figure 2** displays this system via a schematic, a picture and the concurrent circuit diagram. For a more direct comparison between the original and the low-cost system, the same flowmeter was used in both systems, however if implemented the simple 3D-printed stopcock design is able to effectively act as an imprecise flowmeter. Analysis of the electrical output characteristics of this system was performed in the same way as for the system outlined in section 2.1.

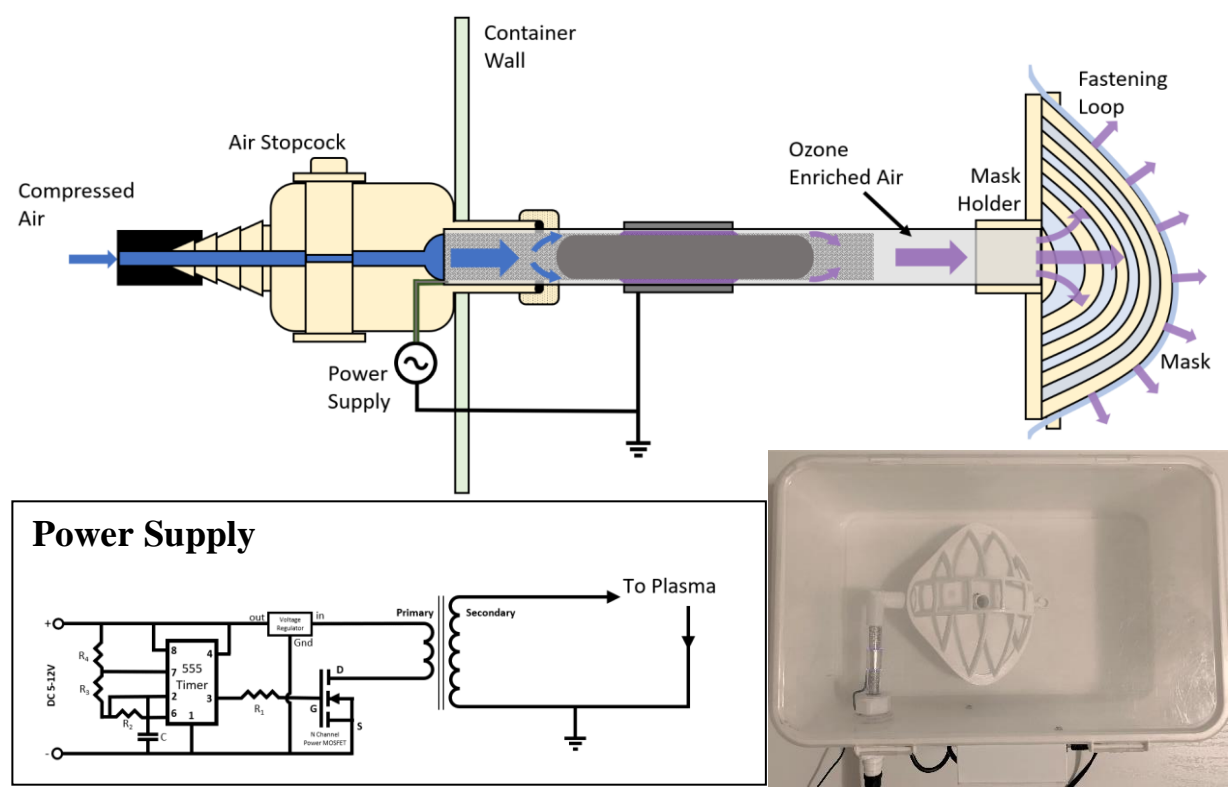

*Figure 2.* Schematic of cost-effective plasma reactor build by using a commercial toy plasma-ball (top), its simplified circuit diagram (bottom left), and a picture of the system (bottom right).

## 2.3 Gas composition characterization

The chemical composition of the gas at the outlet of the plasma discharge was measured via an FTIR spectrometer positioned orthogonally to the plasma stream as shown in **Figure 3**. The reactor was located to one side of a stainless-steel cross connector (KF 25) and, perpendicular to the gas flow, an IR source (Newport 80007) was placed in front of a KBr window. The transmitted light through the gas path was collected after another KBr window by a FTIR spectrometer (Nicolet iS50) (from 800 to 4000 cm$^{-1}$; 50 cumulative averages). Absorption spectra were measured as a function of voltage discharge from 0 to 10 kV. For each applied voltage, a background was acquired before striking the plasma and subtracted.

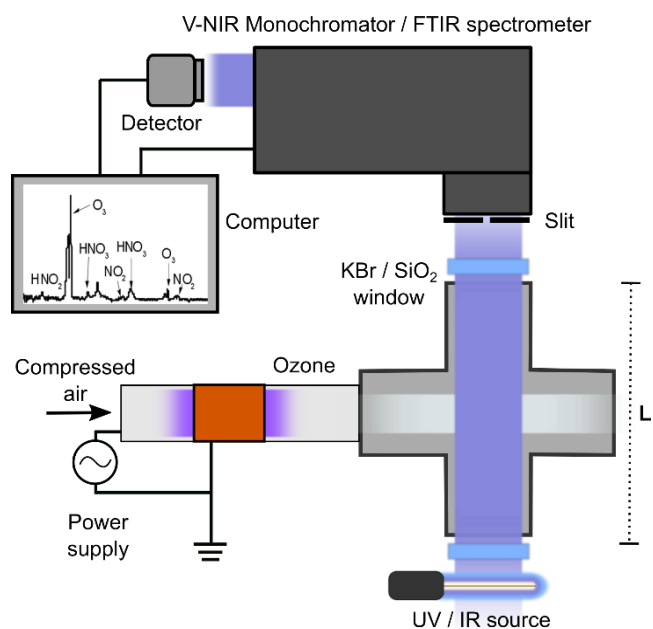

*Figure 3. The same system was used to characterize the chemical composition (FTIR spectrometer and IR source) and the ozone concentration (V-INR monochromator and UV lamp) after the plasma discharge.*

The concentration of ozone after the plasma discharge was carried out by means of UV absorption spectrum with the same configuration used for the chemical composition measurements (see schematic Figure 3). However, a standard UV lamp (Analytik Jen Pen-Ray 90001201) and V-NR spectrometer (Acton Spectra Pro, Princeton Instruments), connected to a CCD camera, substituted the IR source and the FTIR spectrometer, respectively. The KBr windows were also replaced by $SiO_2$ ones to minimize UV light absorption before and after the gas optical path.

The signal intensity at λ=253 nm, where the peak of the ozone absorption cross section is located,[12] was captured by the CCD camera for different discharge voltages from 0 to 10 kV. In addition, the same signal intensity was recorded at different times (from 0 min to 64 min in steps accordingly to the sterilization times) to study the stability of the ozone production. These measurements were used to calculate the ozone concentration by **Equation 1**.[13]

$$C_{oz}[g/m^3] = -\frac{10^6 m_{oz}}{\sigma L N} \log\left(\frac{I_{oz}}{I_0}\right) \quad (1)$$

Where $I_{oz}$ and $I_o$ are the intensity of the signal with and without the presence of ozone respectively, $L$ is the distance in cm of the light path inside the gas (in this case 10.5 cm), σ is the absorption cross section of ozone at approximately $12 \times 10^{-18}$ cm$^2$, $m_{oz}$ is the atomic mass of ozone and $N$ is the Avogadro's number. For consistency with common presentation of ozone concentration as parts per million (ppm), Equation 1 was multiplied by $10^6 \frac{RT}{m_{air}}$, where $R$ is the ideal gas constant, $T$ is the temperature and $m_{air}$ is the atomic mass of air.

## 2.4 Cell culture and inoculation

All assays were performed using *E. coli* β10 cells transformed with a plasmid that confers ampicillin resistance and encodes green fluorescent protein (GFP) under control of a strong constitutive promoter. For each biological replicate, 50 mL of LB media (1% tryptone, 0.5% yeast extract and 1% NaCl) were grown to saturation overnight at 37°C with agitation. Surgical masks were inoculated with 200 µL of culture that were spread on defined 1" x 1" hydrophobic (blue side) regions using sterile scoopulas. Masks were allowed to dry for 60 minutes prior to sterilization.

## 2.5   Quantifying sterilization efficiency

Inoculated segments were excised from masks using sterile scalpels and placed in sterile 50 mL conical tubes. Masks were suspended in 10 mL sterile water and agitated via pulse vortex for 10 seconds. Cells were extracted from masks via centrifugation at 4,000 RPM for 10 minutes. Pelleted cells were resuspended in solution via pulse vortex for 10 seconds. LB agar plates containing 100 μg/mL ampicillin were inoculated using 200 µL of resuspended cellular solution. Each mask was used to inoculate three agar plates as technical triplicates. Agar plates were incubated at 37°C for 16 hours before green fluorescence imaging using a ChemiDoc MP imaging system (Bio-Rad Laboratories, Hercules, CA).

Sterilization kinetics were modelled as the percent of colony forming units (CFUs) relative to control values. Control masks we placed on the mask holder with the device powered off for 16, 32, or 64 minutes. CFUs were counted from fluorescent images using custom MATLAB scripts (MATLAB 2019b; Mathworks, Natick, MA). Control CFUs were calculated as the mean of three technical replicates, and time point measurements are presented as the percent control CFU. Sterilization was modelled as the sum of two exponential decays fit by non-linear least-squares regression in R. Confidence intervals were calculated using a parametric bootstrap with 5000 sample draws.

## 2.6   Mask sterilization and assessment of structural integrity

The effect of the sterilization process on the overall structural integrity of the fibers of the mask was performed with an optical microscope. A medical mask was analyzed with an optical microscope before and after 64 min of ozone treatment (the mask was marked in its center with a sharpie enabling to perform the analysis in a fixed position on the surface of the mask).

## 3   Results and Discussion

The power dissipated in the DBD discharge was characterized following the procedure described by W. Liu et al.[14] The applied voltage $V$ was measured directly from the output of the power supply, while the current flowing through the electrodes was estimated by measuring the charge $Q$ accumulated on a 20 nF measuring capacitor $C_M$ serially connected to the grounded electrode. The Lissajous figure of the DBD discharge was obtained by plotting the measured $Q$-$V$ characteristics (see **Figure 4a**) and the powder dissipated in the discharge was estimated from its area $S$ and discharge frequency $f$ using **Equation 2** (see **Figure 4b**).

$$Power\ (W) = f\ C_M S \qquad (2)$$

The reactor starts coupling a measurable amount of power around 4 kV, linearly increasing with the applied voltage above this threshold (see **Figure 4b**). Fourier Transform Infrared Spectroscopic (FTIR) analysis of the gas produced by the DBD discharge corresponds with the appearance of the typical features of ozone around 4 kV (see **Figure 4c**). In air fed DBDs $O_3$ production is initiated in the plasma phase by the electron impact dissociation of $O_2$ into atomic

O (see **Equation 3**). O quickly reacts with $O_2$ molecules to form $O_3$ via three-body collision (see **Equation 4**, being M a third-body collision partner).[15,16]

$$O_2 + e \rightarrow O + O + e \quad (3)$$

$$O + O_2 + M \rightarrow O_3 + M \quad (4)$$

Other smaller contributions corresponding to $N_xO_y$ species, such as $N_2O_5$ (1250 cm$^{-1}$ and 1720 cm$^{-1}$) and $NO_2$ (1600 cm$^{-1}$ and 1627 cm$^{-1}$), are observed in the spectra.[16] Finally, we observed a sharp feature around 1360 cm$^{-1}$. The sharp was found to be an artifact, attributed to the $O_3$-induced oxidation of the KBr windows. To demonstrate this, we acquired a series of FTIR over few minutes after switching off the plasma (Figure 4d). While the ozone contribution disappears over time, we observe that peak around 1360 cm$^{-1}$ remained unchanged and is hence not related to any gaseous species produced by the plasma discharge.

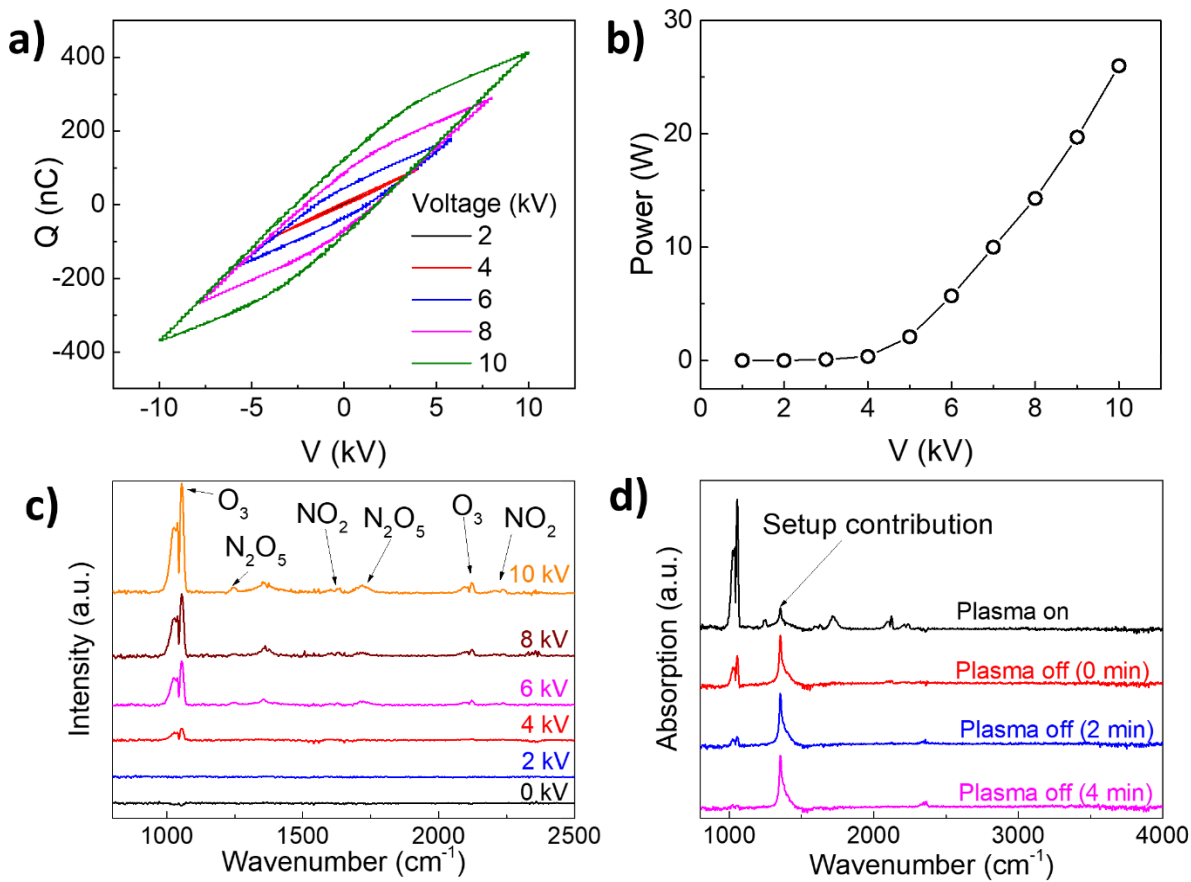

*Figure 4.* (a) Lissajous Figure as a function of applied voltage, (b) coupled power and (c) FTIR measured downstream of the reacator as a functon of applied voltage. (d) FTIR measurement dowstream of the plasma reactor in plasm-on condition, right after switching off the plasa, after 2 min and after 4 min.

The concentration of $O_3$ produced by the DBD discharge as a function of the applied voltage was measured via UV absorption spectroscopy, as described previously in section 2.3. In the first set of experiments the plasma was ignited at a given voltage and allowed to stabilize for 4 minutes before acquiring the measurement. **Figure 5a** shows a linear increase of the gas concentration above 4 kV, reaching a maximum of 750 ppm approximately at 9kV and slowly decreasing above this voltage. This effect has been detailed in the work of S. Yagi et al on air-fed DBD discharges.[17] As the power consumption of the discharge increases, the ozone production shows a correspondent gradual increase, reaches a maximum and then begins decreasing. This effect is likely due to the production of $NO_x$ in the plasma discharge that generates catalytic cycles of $O_3$ destruction (see **Equation 5** and **Equation 6**).

$$O + NO_2 \rightarrow NO + O_2 \tag{5}$$

$$NO + O_3 \rightarrow NO_2 + O_2 \tag{6}$$

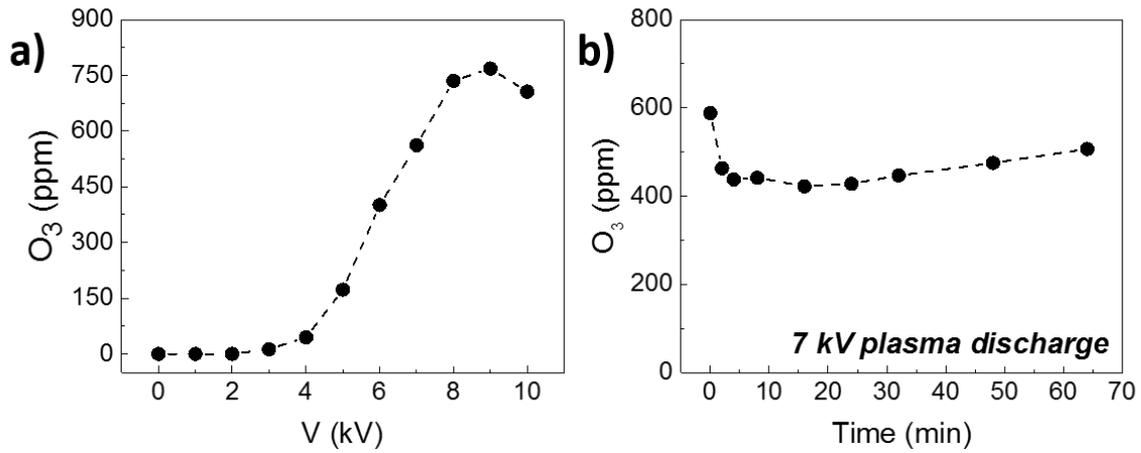

*Figure 5.* a) Ozone concentration as a function of DBD plasma dishcarge voltage. b) Ozone concentration as a function of time ( maximu sterilization time) for 7 kV applied voltage.

For the following sterilization experiments with the DBD system, we fixed the applied voltage at 7 kV as we observed increasing plasma instability above this value. The ozone concentration was measured for 64 for minutes at 7kV, to study the overall stability of the DBD production process. As depicted in **Figure 5b**, after roughly 4 min, the quantity of produced $O_3$ slightly decreases and reaches a stable value.

However, we note a slow concentration increase above 32 minutes, which could be related to thermal stability of the UV lamp itself after remaining on for a long time. On average, the ozone production over 64 minutes is around 453±27 ppm.

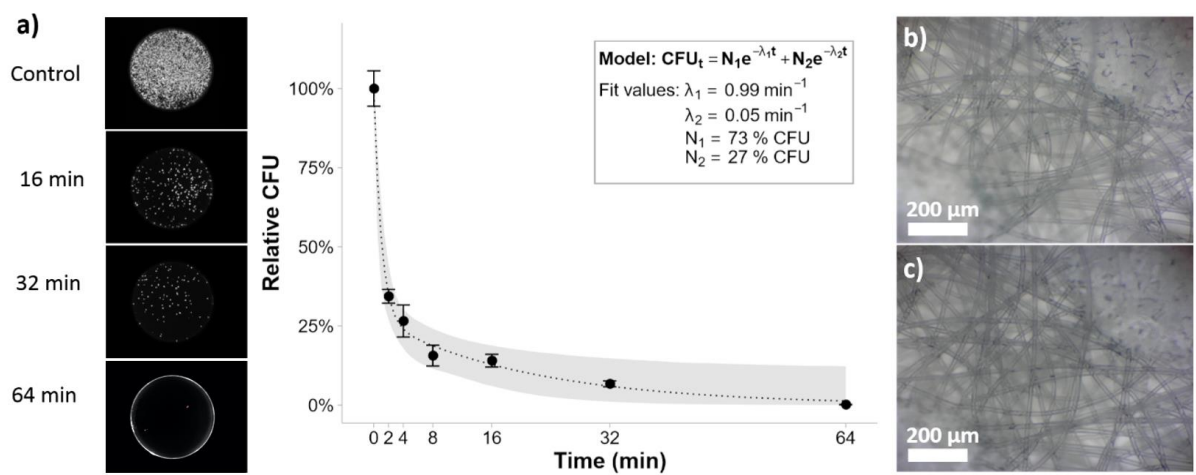

*Figure 6.* (a) Sterilization efficiency as a functon of exposure the $O_3$ time. To the left are fluorscent images of colony growth on agar. The image at 64 minutes is overexposed as only a single colony was observed (*red*). Relative CFU were fit as a sum of two exponential decays. Points represent means and S.E.M. from three biological experiments. The gray ribbon is a 95% confidence interval calculated by a parametric bootstrap. Optical microscope image of a medical mask (b) before and (c) after 64 min of $O_3$ sterlization.

The sterilization efficiency was determined using *E. coli* incubated surgical masks (**Figure 6).** Between 100,000 and 200,000 CFUs were routinely recovered from control masks. CFUs decreased with increasing exposure time. Notably, the change in CFUs exhibits a biphasic behavior that could be modelled as the sum of two exponential curves. This implies 2 populations: A fast dying population with a decay constant $\lambda_1$ (see Figure 6a) of 0.99 min$^{-1}$ and a mean lifetime of approximately 1 minute, and a slow dying population with a decay constant $\lambda_2$ (see Figure 6a) of 0.05 min$^{-1}$ and a mean lifetime of approximately 20 minutes. Across all experiments, we observed a 3:1 ratio ($N_1$:$N_2$, see Figure 6a) between fast and slow dying populations and we speculate that the slow dying population has reduced ozone exposure due to fouling from the saturated bacterial culture. This set of experiments indicated that most bacteria are quickly killed over the first few minutes of the disinfection process, and a complete sterilization is achieved within 64 min (**Figure 6a**). Finally, we performed an analysis of the morphology of the medical mask before and after the sterilization process, to assess any possible structural damage induced by the $O_3$ treatment (**Figure 6b-c**) There was no observable variation of the fiber structure upon sterilization, consistent with the near room temperature operating conditions this method utilizes.

After verifying the effectiveness of the $O_3$ based sterilization in removing bacteria from surgical masks, we explored the characteristics of the "portable-version" of the sterilization apparatus built using the plasma globe toy. The power supply feeds the discharge electrode with a sawtooth signal with frequency equal to roughly 30 kHz and a peak to peak amplitude of 5.6 kV, corresponding to a discharge power of 2 W (**Figure 7a** and **Figure 7b**). Surprisingly, in this configuration we observe an extremely stable $O_3$ production with average concentration, in the order of 1000 ppm (**Figure 7c** and **7d**).

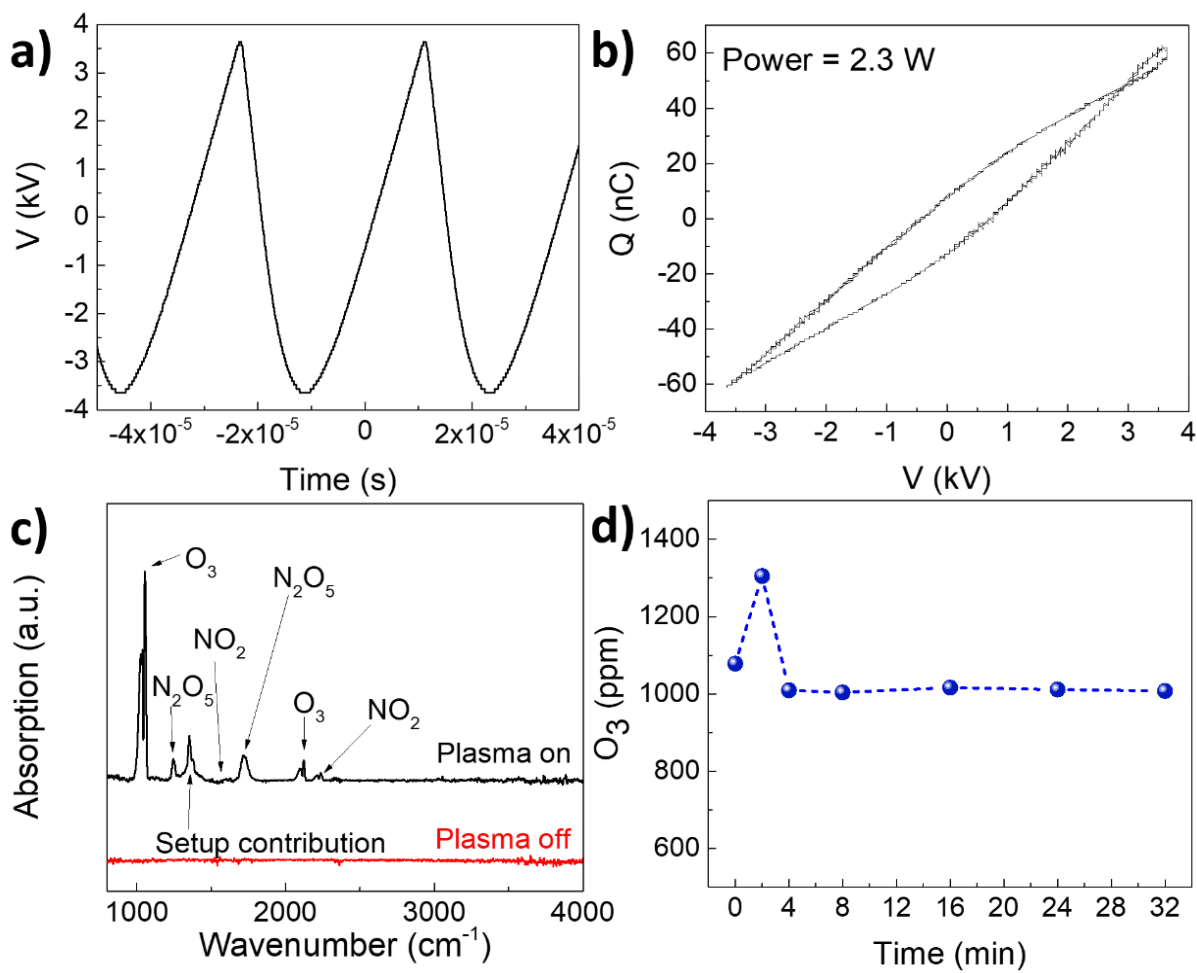

*Figure 7. (a) Voltage signal produced by the power supply of the plasma globe. (b) Lissajous Figure of the plasma globe reactor. (c) FTIR analysis of the gas composition produced by the Plasma Globe Reactor and (c) corresponding $O_3$ concentration produced by the plasma globe reactor as a function of time. An average of 1010 ± 5 ppm along the stability period (4 to 32 minutes).*

Finally, the sterilization efficiency of the plasma globe device was measured for 32 min treatment time and compared with the one of the DBD and the more widespread configuration where the masks is simply placed in a box flooded with $O_3$ (the mask was simply unplugged from the mask holder and placed in the closed 72 L plastic box containing the plasma reactor; the DBD was operated at 7 kV with compressed air flow rate of 10 slm). Correspondingly, we observe a 429% improvement in the sterilization efficiency with respect to the standard configuration. Results summarized in **Figure 8.**

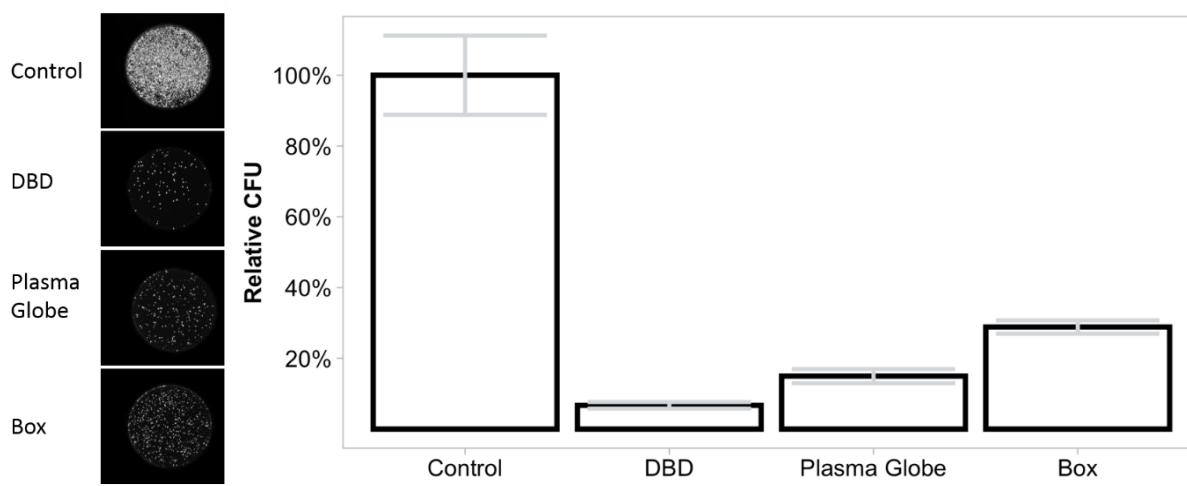

*Figure 8.* Sterilization efficiency at 32 min for the DBD reactor in flow trough configuration, plasma globe reactor and DBD reactor in box configuration. The inset on the left shows fluorscent images of colony growth on agar using different methods of sterilization.

## 4   Conclusion

We have demonstrated that the efficiency of an ozone sterilization system for facepiece respirators can be dramatically increased by careful design of the reactor configurations. Specifically, a flow-through configuration where the ozone is passes directly through the porous fiber structure of the mask demonstrated superior sterilization kinetics with respect to the standard approach of an ozone chamber. Finally, we demonstrate the effective use of a portable single mask sterilization device using low-cost commercially available components (a plasma ball toy, a plastic box, a quartz tube, some steel mesh, HVAC tape, and few 3D printed parts).

Acknowledgements: ((Acknowledgements, general annotations, funding. Normal text.)). We thank Brian Lupish for technical assistance and Mario Leon Lopez for kindly providing the GFP expression plasmid. J.W.C and T.A. are supported by NSF CBET-1951942. J.T.M. is supported by the American Heart Association 19IPLOI34760636. J.S, G.N and L.M. acknowledge support from the U.S. Department of Energy (DOE), Office of Science, Early Career Research Program under award No. DESC0014169. C.B.R. acknowledges the support of the UC Mexus Postdoctoral Fellowship.

Supporting Information ((delete if not applicable))
Additional supporting information is available in the online version of this article at the publisher's website or from the author.
((descriptions of videos or separate figures can be inserted:))
Video S1: ((description - optional))
Figure S1: ((caption - optional))

Graphical Abstract

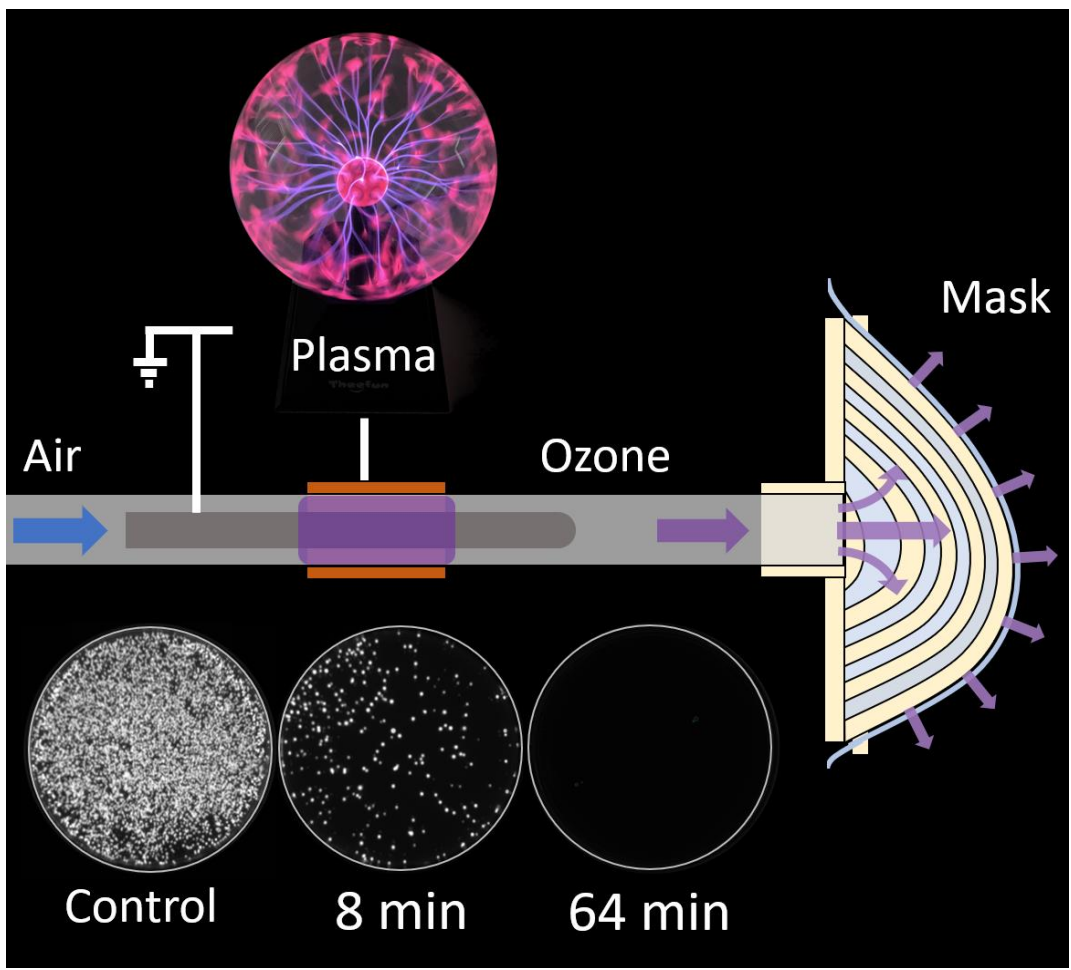

(50mmx55mm so usable for ToC)

**The table of contents entry should be fifty to sixty words long** (max. 400 characters), and the first phrase should be bold. The entry should be written in the present tense and impersonal style. The text should be different from the abstract text.

Respirator sterilization methods are now necessary for medical safety and can be achieved efficiently using flow-through ozone systems powered by materials as affordable as a plasma globe toy. The global COVID-19 pandemic has forced a re-evaluation of sanitization and reuse of what was once single use equipment. This contribution chronicles a simple improvement of ozone-based sanitization of porous media.

C. Author-Two, D. E. F. Author-Three, A. B. Corresponding Author* ((same order as byline))

**Title** ((no stars)): **Efficient Respirator Sterilization via Forced Ozone Convection**

ToC figure ((Please choose one size: 55 mm broad × 50 mm high **or** 110 mm broad × 20 mm high. Please do not use any other dimensions))

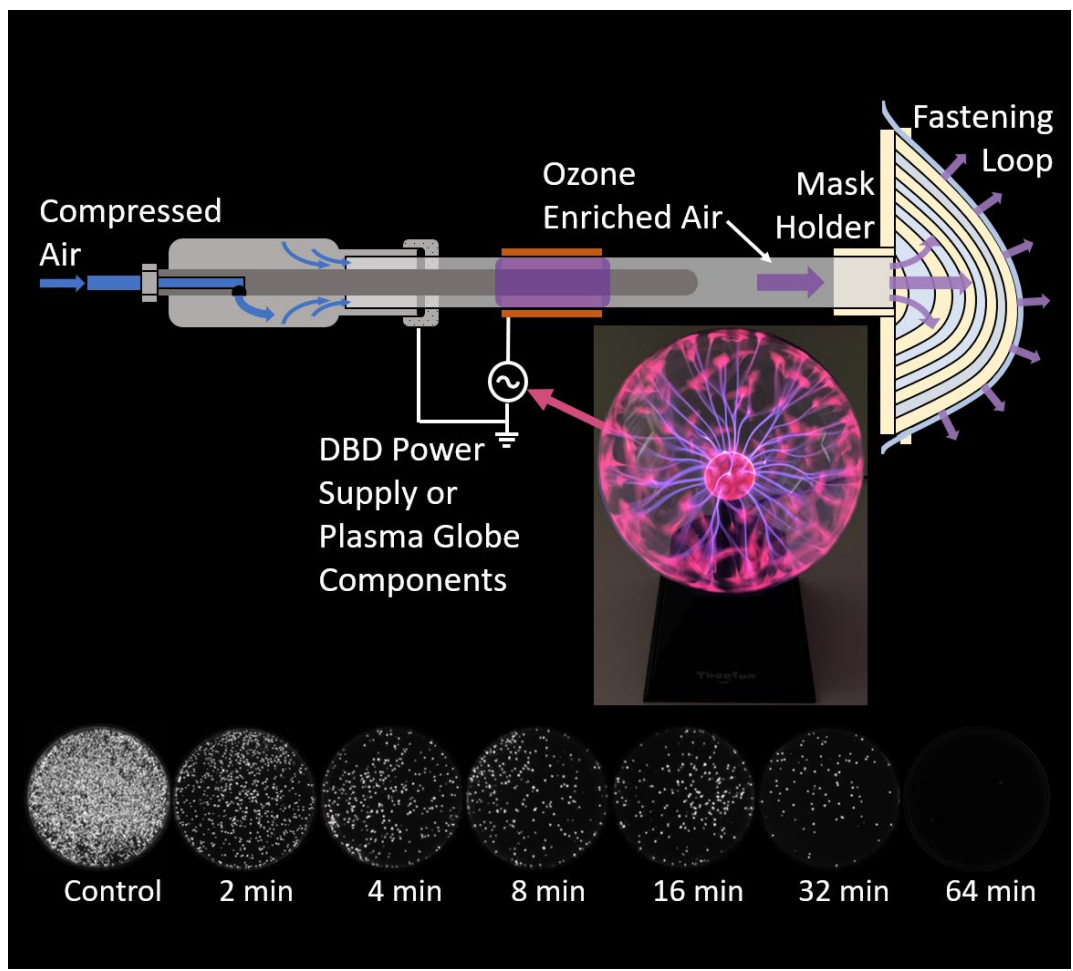

- ((Supporting Information should be uploaded as a separate file))

**Supporting Information**

for *Abbrev. J. Title*, DOI: 10.1002/((please add journal code and manuscript number))

**Title ((no stars))**

Author(s), Corresponding Author(s)* ((write out first and last name))

((Please insert your Supporting Information text/figures here. Please note: Supporting Display items, should be referred to as Figure S1, Equation S2, etc., in the main text…)

CAD Files in STL format: